\documentclass[aps,prx,twocolumn,groupedaddress,longbibliography,superscriptaddress,nofootinbib]{revtex4-1}

\usepackage{graphicx}
\usepackage{amsmath,amssymb,amsfonts}
\usepackage[colorlinks,citecolor=blue,linkcolor=blue,urlcolor=blue, breaklinks=true]{hyperref}
\usepackage{breakurl}
\usepackage{color}
\usepackage[normalem]{ulem}
\usepackage{lineno}
\DeclareMathSymbol{\shortminus}{\mathbin}{AMSa}{"39}

\usepackage[T1]{fontenc}

\usepackage{physics} 
\usepackage[percent]{overpic} 
\usepackage{bm}

\usepackage[english]{babel}

\usepackage{xcolor}
\usepackage{xspace}
\usepackage{ifthen}

\newcommand{\be}{\begin{eqnarray}}
\newcommand{\ee}{\end{eqnarray}}
\usepackage[normalem]{ulem}

\begin{document}
	
	\title{Quantum superposition of black hole states as a communication resource}
	
	\author{Richard Howl}
	\email[Corresponding author: ]{richard.howl@rhul.ac.uk}
	\affiliation{Quantum Group, Department of Computer Science, University of Oxford, Wolfson Building, Parks Road, Oxford,
		OX1 3QD, United Kingdom}
	\affiliation{QICI Quantum Information and Computation Initiative, Department of Computer Science, The University of
		Hong Kong, Pokfulam Road, Hong Kong}
	\affiliation{HKU-Oxford Joint Laboratory for Quantum Information and Computation}
 \affiliation{Department of Physics, Royal Holloway, University of London, Egham, Surrey, TW20 0EX, United Kingdom}
	
	\author{Ali Akil}
	\affiliation{Department of Physics, Southern University of Science and Technology, Shenzhen, 518055, China}
	\affiliation{Department of Physics, The Hong Kong University of Science and Technology, Clear Water Bay, Hong Kong}
 \affiliation{QICI Quantum Information and Computation Initiative, Department of Computer Science, The University of
		Hong Kong, Pokfulam Road, Hong Kong}

	\author{Hl\'{e}r Kristj\'{a}nsson}
	\affiliation{Quantum Group, Department of Computer Science, University of Oxford, Wolfson Building, Parks Road, Oxford,
		OX1 3QD, United Kingdom}
	\affiliation{HKU-Oxford Joint Laboratory for Quantum Information and Computation}
    \affiliation{Perimeter Institute for Theoretical Physics, 31 Caroline Street North, Waterloo, Ontario, N2L 2Y5, Canada}
    \affiliation{Institute for Quantum Computing, University of Waterloo, 200 University Avenue West, Waterloo, Ontario, N2L 3G1, Canada}
    \affiliation{Department of Computer Science and Operations Research, Universit{\' e} de Montr{\' e}al, Montr{\' e}al, Qu{\' e}bec, H3T 1J4, Canada}
	
	\author{Xiaobin Zhao}
	\affiliation{QICI Quantum Information and Computation Initiative, Department of Computer Science, The University of
		Hong Kong, Pokfulam Road, Hong Kong}

	\author{Giulio Chiribella}
	 \email[Corresponding author: ]{giulio@cs.hku.hk}
  \affiliation{QICI Quantum Information and Computation Initiative, Department of Computer Science, The University of
		Hong Kong, Pokfulam Road, Hong Kong}
	\affiliation{Quantum Group, Department of Computer Science, University of Oxford, Wolfson Building, Parks Road, Oxford,
		OX1 3QD, United Kingdom}
	\affiliation{HKU-Oxford Joint Laboratory for Quantum Information and Computation}
	\affiliation{Perimeter Institute for Theoretical Physics, 31 Caroline Street North, Waterloo, Ontario, N2L 2Y5, Canada}
	
    
	\begin{abstract} \noindent 
The interplay between black holes and quantum information has revealed deep connections between gravity and quantum mechanics. For example,  the classical spacetime geometry of a black hole has been shown to degrade quantum entanglement shared by nearby observers, one of whom is in free fall. Here we examine how this picture changes when the black hole’s mass degrees of freedom are allowed to exist in a coherent quantum superposition, corresponding to a superposition of distinct spacetime geometries. We show that such quantum coherence can  mitigate the loss of entanglement between nearby observers, thereby enhancing the performance of quantum communication protocols operating in the vicinity of the black hole.
	\end{abstract}

	\maketitle 
	
	 \section*{Introduction}
	
 Unifying  gravity and quantum mechanics  into a consistent theory  of nature is one of the major unsolved problems in contemporary physics.  Many theories have been proposed, the most prominent of which, string theory and loop quantum gravity, originated several decades ago. Since then, quantum theory has gone through a ``second revolution'' \cite{secondQMRev}, where it was understood that  quantum features, such as coherence and entanglement, can be used as resources to overcome  some of the limitations  of classical physics, leading to technological improvements in areas such as computation and cryptography. This second revolution has led to a new understanding of quantum theory as a theory of information  \cite{Hardy2001,hardy2011reformulating,Hardy_pirsa_09060015,GiulioPurification,Brukner2011,Masanes_2011,barrett2005information,GiulioInformationalDerivation}. It is then natural to explore  whether  information-theoretic tasks could provide a  lens for understanding quantum gravity in operational terms.  	
	
 In this paper, we explore  the implications of  quantum coherence in spacetime degrees of freedom for the transmission of quantum information.   For this purpose, we consider a communication task in the proximity of a black hole, which is an object where quantum gravity is generally expected to play an integral role.  Describing a black hole classically is known to lead to a fundamental decoherence of quantum systems and a degradation of quantum communication protocols that take place near the black hole's event horizon \cite{PhysRevLett.111.021302,AliceFallsIntoBH,BHDeg}. This is connected to other results on how classical gravitational fields can decohere quantum systems, leading to speculation that gravity could be responsible for our world appearing classical at macroscopic scales \cite{PhysRevLett.111.021302,PhysRevD.93.044027,Anastopoulos_2013,pikovski2015universal,Bassi_2017,GravInQuantumLab}, see also \cite{DIOSI1987377,penrose1996gravity,Howl_2019} for earlier work on gravity  addressing the measurement problem of quantum theory. All these works considered a classical theory of gravity.

We now explore the possibility that a black hole  can be described by standard quantum theory. Specifically, we consider the information–theoretic consequences of two assumptions that one may expect to hold in a fully-fledged theory of quantum gravity: (i) a black hole can be in a coherent superposition of  classical mass states,  entangled with an auxiliary system, and (ii) it is possible to perform a measurement  on the auxiliary system to probe the quantum interference between the classical mass states. Later in the paper we provide several arguments for how these assumptions might be satisfied. We stress, however,  that the aim of the paper is not to argue for the validity of these assumptions, but rather to  provide a thought experiment exploring  their information–theoretic consequences. Our main result is that quantum coherence in the black hole spacetime, if present, can allay the (previously considered fundamental) decoherence of quantum systems, thereby improving the performance of quantum communication protocols performed in the proximity of a black hole.

\section*{Results}
	 
\subsection*{Quantum communication near classical black holes} 
	 
Here, we briefly review  quantum communication near a classical black hole \cite{BHDeg,AliceFallsIntoBH}. This is also analogous to quantum communication when one or two of the parties are  accelerating in flat spacetime \cite{PhysRevLett.91.180404,PhysRevA.74.032326,AliceFallsIntoBH}, which  played a pivotal role in the development of relativistic  quantum information \cite{RevModPhys.76.93,RQIReview,Alsing_2012,GravInQuantumLab}. For a more detailed review, see Appendix \ref{app:review}.

Consider two parties, Alice and Rob that both have access to  a scalar fermionic (Grassmann) quantum field near the event horizon of a Schwarzschild black hole. Alice and Rob  share a  bipartite quantum state of this field that is  maximally entangled for observers in free fall:
 \begin{align} \label{eq:psiAR}
		|\psi \rangle_{AR} =\left(|0_{\bm{k}_A}\rangle^+_A |0_{\bm{k}_R}\rangle^+_R +   |1_{\bm{k}_A} \rangle^+_A |1^{\mathfrak{R}}_{\bm{k}_R}\rangle_R^+ \right) / \sqrt{2}.
\end{align}
Here, $|0_{\bm{k}_A,\bm{k}_R}\rangle^+_{A,R}$ represents the Minkowski  vacuum state in field modes $\bm{k}_A,\bm{k}_R$ for Alice and Rob, $|1_{\bm{k}_A}\rangle^+_A $ is a one-particle Minkowski state for Alice in mode $\bm{k}_A$, and $|1^{\mathfrak{R}}_{\bm{k}_R}\rangle_R^+$ is a one-particle right Unruh state for Rob in mode $\bm{k}_R$. All other modes are assumed to be in vacua. Unruh modes span the  Minkowski modes and   have the same  vacuum $|0\rangle_M$ as Minkowski modes \cite{BHDeg}. 

\begin{figure}[t]
		\includegraphics[width=8cm]{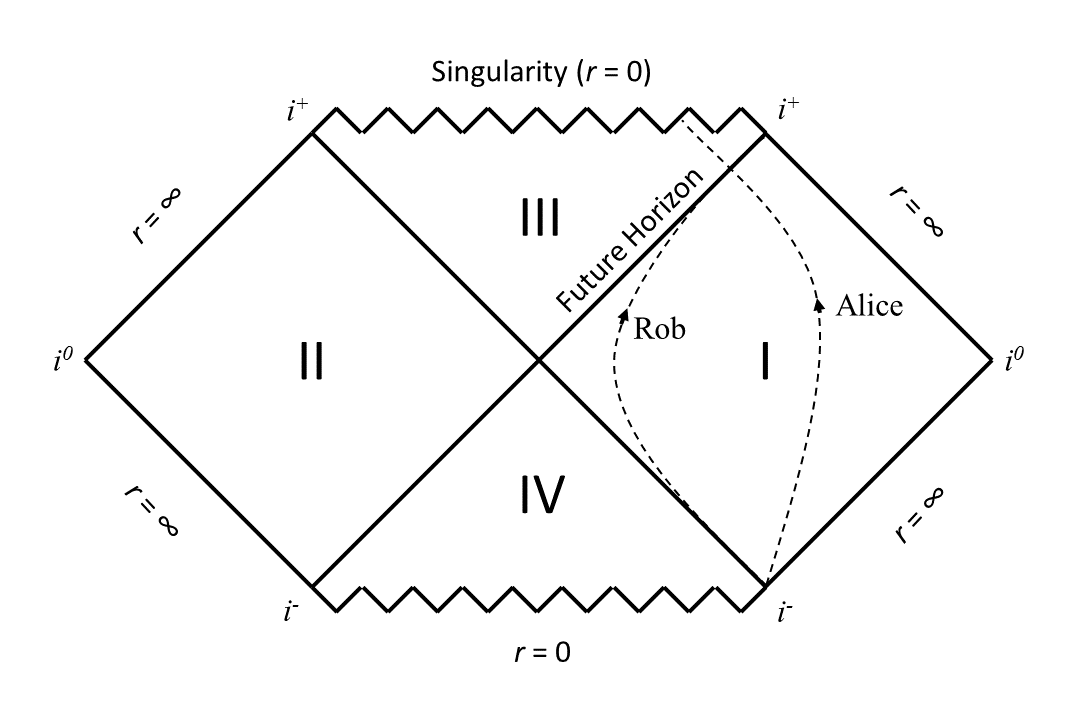}
		\caption{Penrose diagram of a Schwarzchild black hole. Rob is stationary above the event horizon and constrained to region I. Alice is free-falling into the black hole. Rob in region I is causally disconnected from region II. Alice and Rob are assumed to share an entangled state of a scalar fermionic field, which could be used e.g. for a quantum communication protocol. For example, as considered in the main text, Alice could attempt to teleport a qubit state $|\Psi \rangle = \alpha |0_{\bm{k}_A} \rangle^+ +   \beta |1_{\bm{k}_A} \rangle^+$ to Rob. The teleportation protocol can take place at any time before Alice crosses the horizon.} 
		\label{fig:PenDiag}
	\end{figure}

	We now consider that Alice is free-falling just above  the event horizon of a black hole, while Rob is stationary a small distance above the horizon. We follow \cite{BHDeg} in considering Alice to be free falling just above the event horizon, and assume that she does not experience Hawking radiation. Alternatively, we could also consider Alice to be free falling but very far from the horizon, in which case the Hawking radiation she would experience would be negligible.	Alice, being a free-fall observer, will define her vacuum and excited states to be that of Minkowski space. Rob, however, will define his vacuum to be the Rindler (or equivalently Boulware) vacuum \cite{BHDeg}. This results in the following transformation to the state $|\psi \rangle_{AR}$:
	\begin{align} \label{eq:xi}
		\xi(\hat{\rho}_{\psi}) = \mathrm{Tr}_{\shortminus \bm{k}_R} \left[\hat{U}_{AR} (r) \left( \hat{\rho}_{\psi} \otimes \hat{\rho}_{\shortminus \bm{k}_R}  \right)\hat{U}^{\dagger}_{AR} (r) \right], 
	\end{align}
	where $\rho_{\psi} := |\psi \rangle_{AR} \,_{AR}{\langle} \psi |$; $\hat{U}_{AR}(r) := I_A \otimes \hat{U}_{\bm{k}_R} (r)$, $\hat{\rho}_{\shortminus \bm{k}_R} := |0_{\shortminus \bm{k}_R} \rangle^-\, ^{-}{\langle} 0_{\shortminus \bm{k}_R} |$, with  $\{+,-\}$ indicating particle and antiparticle states; and $\mathrm{Tr}_{\shortminus \bm{k}_R}$ means that we trace over mode $\shortminus \bm{k}_R$ in regions I and II of the black hole spacetime. The spacetime of a black hole can be represented by a Penrose diagram (see Figure \ref{fig:PenDiag}), consisting of four regions, I, II, III and IV, with regions I and II being outside the black hole and regions III and IV representing being beyond the respective event horizons of each region. Near the event horizon, regions I and II play the role of the right and left wedges $\mathfrak{R}$ and $\mathfrak{L}$ of Rindler space \cite{thorne2000gravitation,Fabbri:2005mw,BHDeg}.  We trace out the Rindler modes of II in \eqref{eq:xi} because Rob is assumed in region I of the black hole spacetime, which is casually disconnected from  II.
	
	Above, $\hat{U}_{\bm{k}_R} (r)$ describes the change of basis from the Unruh modes of Minkowski space to Rindler modes of Rindler space. It is a two-mode squeezing unitary 
	\cite{PhysRevA.74.032326}:
	\begin{align} \label{eq:Uk}
		\hat{U}_{\bm{k}} (r)  = e^{r (\hat{c}^{\mathfrak{R}}_{\bm{k}} \hat{d}^{\mathfrak{L}}_{\shortminus \bm{k}} e^{i \phi} + \hat{c}^{\mathfrak{R} \dagger}_{\bm{k}} \hat{d}^{\mathfrak{L} \dagger}_{\shortminus \bm{k}} e^{-i \phi})},
	\end{align}
	such that $\bm{B}_{\bm{k}} := \hat{U}_{\bm{k}} \bm{C}_{\bm{k}} \hat{U}^{\dagger}_{\bm{k}}$, where $\bm{B}_{\bm{k}} = (\hat{a}_{\bm{k}}, \hat{b}^{ \dagger}_{\shortminus \bm{k}})^T$, $\bm{C}_{\bm{k}} = (\hat{c}^{\mathfrak{R}}_{\bm{k}}, \hat{d}^{\mathfrak{L} \dagger}_{\shortminus \bm{k}})^T$, and $\phi$ is a phase that is often neglected on the grounds of being physically irrelevant \cite{PhysRevA.74.032326,Audretsch1994} (it is possible that with a quantum theory of spacetime curvature,  there is a physical interpretation of the phase, and in Appendix \ref{app:SupPhases}, we show that, if it were physical, the phase can also be used to allay degradation of entanglement).  Here,  $\hat{a}^{\mathfrak{R}}_{\bm{k}}$ is the right Unruh particle annihilation operator for  mode $\bm{k}$, $\hat{b}^{\mathfrak{R}}_{\bm{k}}$ is the right Unruh anti-particle annihilation operator, $\hat{c}^{\mathfrak{R}}_{\bm{k}}$ is the  annihilation Rindler particle operator  in region $\mathfrak{R}$,   $\hat{d}^{\mathfrak{L}}_{\bm{k}}$ is the  annihilation Rindler anti-particle operator in region $\mathfrak{L}$, and $r$ is related to the acceleration of the Rindler observer. Taking the observer to be near the event horizon of a black hole and using the equivalence of the Rindler and Schwarzchild spacetimes in this region, $r$ is  found through \cite{Audretsch1994,BHDeg}:
	\begin{align} \label{eq:tanr}
		\tan r = e^{- \hbar \pi \sqrt{f_0} k^0 / \kappa}, 
	\end{align}
	with $f_0 := 1 - 2m / R_0$; $k^0$ the frequency of Rob's mode; $\kappa := 1 / (4m)$, where $m$ is the mass of the black hole, and $R_0$   is Rob's position in the Schwarzschild radial coordinate. We are working in units $G=c=1$, and so $2 m$ is also the radius $R_s$ of the event horizon of the black hole. The requirement that Rob is close to the event horizon is expressed by $(R_0 - R_s) \ll R_s$. Notice, moreover, that since the exponent on the right-hand side is negative, the squeezing parameter $r$ ranges from 0 to $\pi/4$.
    
	The transformation \eqref{eq:xi} is analogous to a  fermionic, Gaussian-amplification quantum channel, which can be decomposed in terms of Kraus operators:  $\xi(\hat{\rho}_{\psi}) = \sum_n \hat{M}_n \hat{\rho}_{\psi} \hat{M}^{\dagger}_n$, with $\sum_n \hat{M}^{\dagger}_n \hat{M}_n = I$. There are only two such Kraus operators \cite{banerjee2017characterization}, which are presented in Section \ref{app:Kraus}. After the action of the quantum channel, the maximally-entangled state $\rho_{\psi}$ becomes \cite{PhysRevA.74.032326}:
	\begin{align} \label{eq:xirho}
		\xi(\hat \rho_{\psi}) &= \frac{1}{2} \left(\begin{array}{cccc} \cos^2 r & 0 & 0 & \cos r\\ 0 & \sin^2 r & 0 & 0 \\ 0 & 0 & 0 & 0 \\ \cos r & 0 & 0 & 1 \end{array} \right),
	\end{align}
	with the state in the Hilbert space $\mathcal{H}_A \otimes \mathcal{H}^I_R$, where the former  is Alice's Hilbert space and the latter is Rob's  Hilbert space over  region I.	Measuring the entanglement of this state using the negativity measure $\mathcal{N}[\rho] := (|| \rho^{\Gamma_A}||_1 - 1)/2$, the entanglement is found to be $\mathcal{N} [\xi(\rho_{\psi})] = \cos^2 r / 2$ \cite{PhysRevA.74.032326}. 
	
	The inertial maximally entangled state \eqref{eq:psiAR} has therefore become a non-maximally entangled mixed state for Rob standing next to the event horizon of a black hole.  The entanglement resource is thus degraded and this will affect quantum communication between the two parties. For example, if Alice wants to teleport some state $|\Psi \rangle = \alpha |0_{k_A} \rangle^+ +   \beta |1_{k_A} \rangle^+$ to Rob, with $\alpha$ and $\beta$ any complex numbers that normalize the state,  then the best we can expect Rob to recover at the end of the protocol is $|\Psi \rangle = \alpha |0^I_{\bm{k}_R} \rangle^+ +   \beta |1^I_{\bm{k}_R} \rangle^+$, where $|0^I_{\bm{k}_R} \rangle^+$ and $|1^I_{\bm{k}_R} \rangle^+$ are the vacuum and single particle states of region I  \cite{PhysRevLett.91.180404,PhysRevA.74.032326}. If Rob is in free fall like Alice, then this protocol is perfect when using their shared maximally entangled state $\rho_{\psi}$. However, if Rob sits just above the event horizon of the black hole, then the fidelity of the protocol necessarily decreases since the entanglement of $\rho_{\psi}$ is degraded \cite{PhysRevLett.91.180404,PhysRevA.74.032326,BHDeg}.

	\subsection*{Using quantum gravity as a resource}
	
As reviewed in the previous section, the presence of a black hole  causes an inherent degradation of entanglement and thus also quantum communication protocols taking place near the black hole \cite{BHDeg,AliceFallsIntoBH,PhysRevLett.91.180404,PhysRevA.74.032326}. This together with its flat spacetime analogue with accelerated observers  played a central role in shaping the field of relativistic quantum information \cite{RevModPhys.76.93,Alsing_2012,RQIReview}. However, its derivation assumed a classical description of a black hole and its spacetime. Would the same conclusions hold if we were to treat the black hole's spacetime quantum mechanically? This question is important because most theories of quantum gravity predict quantum effects to play an integral role in black holes. Here we show that the quantum structure of spacetime can be used as a resource to allay the degradation of quantum entanglement in the proximity of purple the black hole.

\begin{figure}[t]
		\includegraphics[scale=0.35]{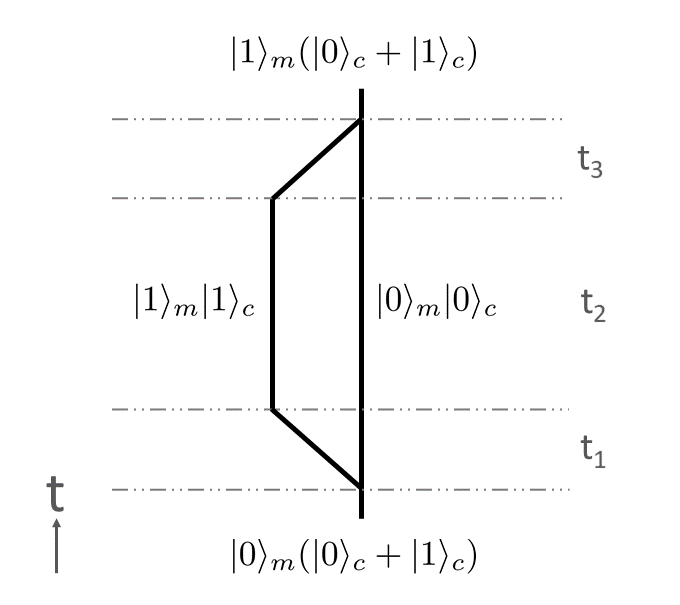}
		\caption{Quantum communication protocol using a quantum black hole. The black line indicates the black hole state. Before time period $t_1$, the black hole is in a single mass state $\ket{0}_m$, and the control is in the superposition state $(\ket{0}_c + \ket{1}_c)/\sqrt{2}$. The black hole then moves into an entangled state with the control: if the control is $\ket{0}_c$ the black hole remains in mass state $\ket{0}_c$, whereas if the control is $\ket{1}_c$, the black hole is placed in mass state $\ket{1}_m$. During this time, Alice and Rob share an entangled state. Rob then moves his state to a higher energy mode and accesses the control - the black hole is then subsequently always placed in the state $\ket{1}_m$ irrespective of the state of the control. Finally, Rob measures the control and then performs a quantum communication with the shared entangled state protocol. For example, he could perform quantum teleportation to teleport Alice's state.}
		\label{fig:BHSuperposition}
	\end{figure}

Our first assumption is that a black hole can be in a superposition of masses $m_1$ and $m_2$,  entangled with an auxiliary control  system $c$.
 For example, Rob could hold a quantum machine that accesses a control qubit such that if the control is in state $|0\rangle_c$ it does nothing, while if $|1\rangle_c$ it drops a mass $m = m_2 - m_1$ into the black hole - see Figure \ref{fig:BHSuperposition}. A superposition of masses of black holes can also occur naturally due to black hole evaporation (see Appendix \ref{app:BHSup}). However, for simplicity we ignore black hole evaporation by either assuming a supermassive  black hole or that the black hole is in equilibrium such that the amount of energy lost due to evaporation is compensated by the energy from infalling objects.
	
	With the black hole in a superposition of two masses, it is natural to expect its gravitational field, and therefore the structure of spacetime, to also be in a quantum superposition, resulting in a superposition of spacetime event horizons.  Assuming that Alice is always in free fall, and  Rob is always standing outside the event horizon at a fixed $R_0$ (his rocket  automatically adjusts to the gravitational field it is in), Alice and Rob's state $|\psi\rangle_{AR}$ of \eqref{eq:psiAR} would undergo a superposition of two channels with squeezing factors $r_1$ and $r_2$ defined by \eqref{eq:tanr} with $m$ replaced by $m_1$ and $m_2$ respectively. Note that in a Schawrzschild geometry, the Schwarzschild radial coordinate $r$ coincides with the areal radius, i.e.\ the radius of a sphere enclosing an area $4 \pi r$. Moreover, in the Schawrzschild geometry, this areal radius does not depend on the black hole mass. Therefore, the definition of $R_0$, where Rob stands, is the same in all branches of the superposition. Note also that Rob's  rocket will become entangled with the black hole, and thus Rob's rocket can play the role of the control qubit. 
    
    The state of Alice's and Rob's modes thus becomes entangled with the black hole and control:
    \begin{align} \nonumber
	 	|\Psi\rangle_{A R m c}  := \frac{1}{\sqrt{2}} \Big( &\hat{U}_{AR} (r_1)  |0 \rangle_m |0\rangle_c \\\label{eq:SupChannels}&+ \hat{U}_{AR} (r_2)  |1 \rangle_m |1\rangle_c\Big) |\psi\rangle^0_{AR},
	 \end{align}
	where  $|0 \rangle_m$ and  $|1 \rangle_m$ respectively denote the two mass states of the black hole with masses $m_1$ and $m_2$, and $|\psi\rangle^0_{AR}:= |\psi\rangle_{AR} \otimes |0_{\shortminus \bm{k}_R}\rangle^-$.   Here, for simplicity, we have chosen the relative phase between the two unitaries $\hat{U}_{AR} (r_1)$ and $\hat{U}_{AR} (r_2)$ to be zero. We have also ignored all other modes $\bm{k}\neq \bm{k}_A, \bm{k}_R$ for simplicity as they are not relevant to the protocol outlined below. Tracing over mode $\shortminus \bm{k}_R$, we  have:
	\begin{align}\nonumber
		\hat{\rho}_{ARmc} = \frac{1}{2} &\Big[	\xi_{11}(\hat{\rho}_{\psi}) \otimes |0 0\rangle_{mc}  \,_{mc}{\langle} 00| \\\nonumber&+  \xi_{22}(\hat{\rho}_{\psi}) \otimes |11 \rangle_{mc} \,_{mc}{\langle} 11|\\ \label{eq:FullChannel} &\nonumber+\xi_{12}(\hat{\rho}_{\psi}) \otimes |00 \rangle_{mc} \,_{mc}{\langle} 11| \\&+  \xi_{21}(\hat{\rho}_{\psi}) \otimes |11 \rangle_{mc} \,_{mc}{\langle} 00| \Big],
	\end{align}
	where 
	\begin{align}\nonumber
		\xi_{ij}(\hat{\rho}_{\psi}) &:= \mathrm{Tr}_{\shortminus \bm{k}_R} \left[\hat{U}_{AR} (r_i) \left( \hat{\rho}_{\psi} \otimes \hat{\rho}_{\shortminus \bm{k}_R} \right)\hat{U}^{\dagger}_{AR} (r_j) \right],
	\end{align}
	with $i,j \in \{1,2\}$.

Our second assumption is that Rob is able to perform a measurement that witnesses the interference between the quantum channels arising from the $|0\rangle$ and  $|1 \rangle$ mass states. The physical implementation of such a measurement, of course, is difficult to describe in the lack of an established theory of quantum gravity. Nevertheless, the in-principle possibility of measuring quantum observables on a black hole seems to be a necessary feature of any theory that treats the black hole as a quantum object. Na\"ively, one might expect that this would require a projective measurement on the black hole and control in the basis $\{|+\rangle_m,|-\rangle_m\}$ and $\{|+\rangle_c,|-\rangle_c\}$,  where $|+\rangle := (|0\rangle + |1 \rangle)/\sqrt{2}$ and $|-\rangle := (|0\rangle - |1 \rangle)/\sqrt{2}$. However, we show that such exotic measurements need not be necessary. 
Some candidate mechanisms are discussed in the following. 

First, if the masses $m_1$ and $m_2$ are close in value and we assume coherent states, a parity measurement would approximately implement the desired measurement, as discussed in Appendix \ref{app:BhPVM}. Alternatively, \emph{we can avoid a quantum measurement on the black hole entirely} by assuming that Rob's machine shifts his state to a higher energy mode, and then accesses the control again. This time, if  the control is in state $|0\rangle_c$ the  machine drops a mass $m$ into the black hole, whereas if it is in $|1\rangle_c$ the machine does nothing. The reduced state $\hat{\rho}_{ARmc}$ is then $ |1\rangle_m \, _{m}{\langle} 1| \otimes \hat{\rho}_{ARc}$, where $\hat{\rho}_{ARc}$ is
	\begin{align}\nonumber
		\hat{\rho}_{ARc} &= \frac{1}{2}  \Big[	\xi_{11}(\hat{\rho}_{\psi}) \otimes | 0\rangle_{c} \, _{c}{\langle} 0| \\\nonumber&+  \xi_{22}(\hat{\rho}_{\psi}) \otimes |1 \rangle_{c} \, _{c}{\langle} 1|+\xi_{12}(\hat{\rho}_{\psi}) \otimes |0 \rangle_{c} \, _{c}{\langle} 1| \\ \label{eq:rhoARc} &+  \xi_{21}(\hat{\rho}_{\psi}) \otimes |1 \rangle_{c} \, _{c}{\langle} 0| \Big],
	\end{align} 	
	
	Rob now measures the control in the $\{+,-\}$ basis. The {\em a posteriori} joint states of Alice and Rob  would then be $\hat{\rho}^+_{AR} := \,_{c}{\langle} +  |  \hat{\rho}_{ARc} | + \rangle_{c} / \mathrm{Tr} [\,_{c}{\langle} + |  \hat{\rho}_{ARc}  |+ \rangle_{c}]$ and $\hat{\rho}^-_{AR} := \,_{c}{\langle}-|  \hat{\rho}_{ARc}  |-\rangle_{c} / \mathrm{Tr} [\,_{c}{\langle}-|  \hat{\rho}_{ARc}  |-\rangle_{c}]$, as shown in Section \ref{app:SupChannels}. 	It is found that  $\hat{\rho}^-_{AR}$ is a separable state such that it is now possible for the entanglement between Alice and Rob's systems to completely decohere, and this is independent of the values of $r_1$, $r_2$, $\phi_1$ and $\phi_2$. This is in stark contrast to the classical case considered above, where there is always entanglement between Alice and Rob's systems.
	
\begin{figure}[t]
		\includegraphics[trim={1.2cm 0 0 0},clip,width=7cm]{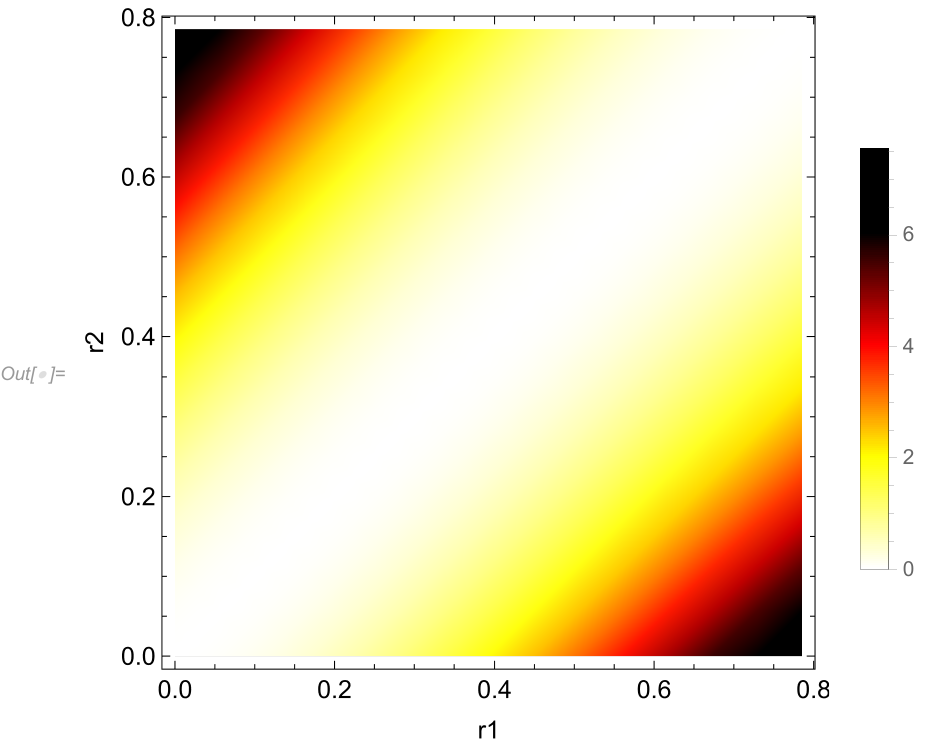}
		\caption{Density plot of percentage difference between the average entanglement negativity and a classical mixture. This is provided for the squeezing factors   $r_1$ and $r_2$ of the two channels for the physical range  $0$ to $\pi/4$. White through yellow and red to black indicates small to larger percentage differences. The white area corresponds to the case where the masses in superposition are close in value, resulting in the  classical case when the masses are exactly the same.}  \label{fig:PerDiff}
	\end{figure}

		The average entanglement that Rob would obtain after measuring the control and keeping the results of his measurements is, after many measurements, $p_+ \mathcal{N}[\rho_+]$, where  $p_+$ is the probability that Rob measures $|+\rangle_c$. This is found to be 
	\begin{align}\nonumber
		\mathcal{N}_{Av}[\rho] &= \frac{1}{16} \Big( -(\sin r_1 + \sin r_2)^2 \\&+ \sqrt{16(\cos r_1 + \cos r_2)^2 + (\sin r_1 + \sin r_2)^4}\Big).
	\end{align} 
	This is to be compared to the entanglement obtained if there were instead a classical mixture of black holes with masses $m_1$ and $m_2$, which is the same as that obtained when tracing out the control:
	\begin{align}\nonumber
		\mathcal{N}[\rho_{Av}] &= \frac{1}{8} \Big(-(\sin^2 r_1 + \sin^2 r_2) \\&+ \sqrt{4(\cos r_1 + \cos r_2)^2 +  (\sin^2 r_1 + \sin^2 r_2)^2}\Big),
	\end{align}
	where $\rho_{Av} := [	\xi_{11}(\hat{\rho}_{\psi})  +  \xi_{22}(\hat{\rho}_{\psi})]/2$. 	The average entanglement obtained from the quantum superposition of channels $\mathcal{N}_{Av}[\rho]$ is found to always be greater than $\mathcal{N}[\rho_{Av}]$, which is illustrated in Figure \ref{fig:PerDiff}. This occurs despite $\rho_-$ being separable as the chance of obtaining this state is always extremely  low.  

In summary, quantum coherence in spacetime degrees of freedom  can in principle be used as a resource to allay the degradation of entanglement brought about by a black hole.    This  then improves the fidelity of quantum teleportation near a black hole \cite{PhysRevLett.91.180404,PhysRevA.74.032326} as discussed above, and, as we will see in the following, can increase the communication capacity of the noisy channel acting on Rob's mode.   Note that equivalently we could have obtained the same result considering, for instance, a superposition of accelerations (as would be predicted by the equivalence principle) or a superposition of distances from the black hole.
	
	\subsubsection*{Quantum capacity of communication channel} \label{app:CohInfo}

	Above we considered how quantum gravity can be used to allay degradation of entanglement between Alice and Rob. For quantum communication, the ability of Alice and Rob to communicate a qubit between each other is quantified by the quantum capacity of the communication channel between them. The quantum capacity of any channel is lower bounded by the coherent information of the channel, which in turn is lower bounded by the coherent information of the output state resulting from the action of the channel on one half of the maximally entangled state. In the scenario considered above, the effective channel described by \eqref{eq:rhoARc} (arising from the superposition of two fermionic Gaussian amplification channels with squeezing factors $r_1, r_2$, respectively) can be thought of as a communication channel between Alice and Rob, where half of the maximally entangled state is sent from Alice to Rob, with Alice keeping the other half herself. Therefore, the coherent information of the output state of \eqref{eq:rhoARc} gives a lower bound to the quantum capacity of the communication channel available between Alice and Rob. 
	
		\begin{figure}[t]
		\includegraphics[width=8cm]{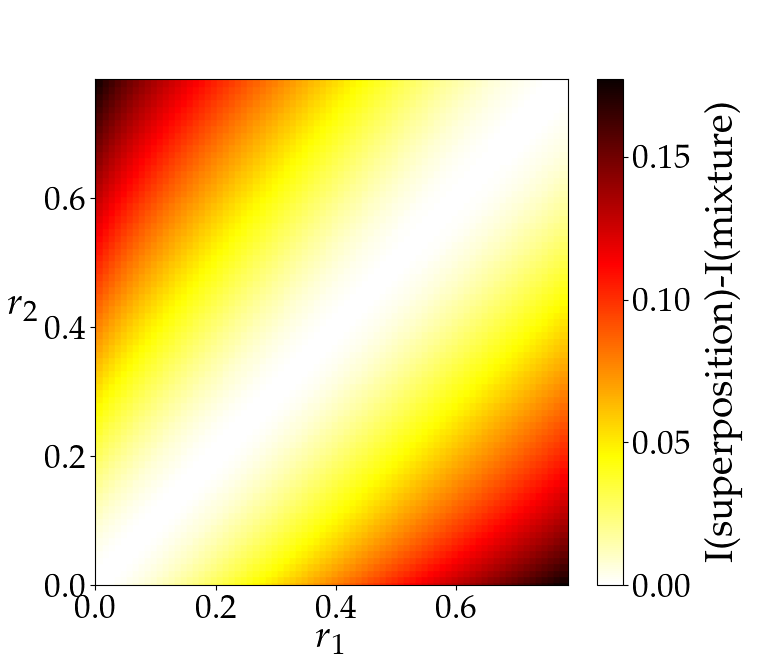}
		\caption{Density plot of the difference between the coherent information of a superposition of Hawking channels and their classical mixture. This is provided for the squeezing factors   $r_1$ and $r_2$ of the two channels for the physical range  $0$ to $\pi/4$. White through yellow and red to black indicates small to larger differences.} \label{fig:CoherentInfo}
	\end{figure}
	
	Similarly, one can calculate the coherent information of the output state arising from a classical mixture of  two fermionic Gaussian amplification channels with squeezing factors $r_1, r_2$. The difference in the coherent information between the superposition case and the classical mixture case can be used to quantify the effect of a superposition of black hole masses on quantum communication.
	
	The coherent information of a bipartite quantum state $\hat{\rho}_{AB}$ is given by $I(A,B)_\rho = S(B )_\rho - S(A B )_\rho$, where $S(AB)_\rho$ is the von Neumann entropy of $\hat{\rho}_{AB}$ and $S(B)_\rho$ is the Von Neumann entropy of the reduced state $\hat{\rho}_B := \Tr_A \hat{\rho}_{AB}$. 
	
	Figure \ref{fig:CoherentInfo} provides a density plot of the difference in the coherent information between the output state of the effective channel \eqref{eq:rhoARc}, arising from a superposition of two channels, and the corresponding output state for a classical mixture of the two channels, for $r_1$ and $r_2$ between $0$ and $\pi/4$. We find that this difference is always positive, indicating that the superposition of channels can increase the ability to communicate quantum information between the two relativistic agents.
	
\subsubsection*{$N$ superposition states}

We have thus seen that an entangled state near a black hole decoheres less if the black hole is in a superposition of two masses compared to being in a classical mixture of two masses. It is then reasonable to ask whether having a superposition of $N>2$ states  would decrease the state decoherence even further. If a larger $N$ gives a smaller decoherence, one might also expect that for large enough $N$ there would potentially be no decoherence at all. This can also be motivated by the results in \cite{Hler2019}, showing that a superposition of $N$ independent and identical decohering quantum channels for large $N$ can interfere to produce a perfect communication channel for quantum information. However, one important difference between \cite{Hler2019} and the case considered in this work is that the former results there were derived for the case where the decohering channels were independent, while in our case the channels corresponding to different mass states are correlated, as they arise from tracing out the \textit{same} state of the environment.

\begin{figure}
    \centering
\includegraphics[width=0.9\linewidth]{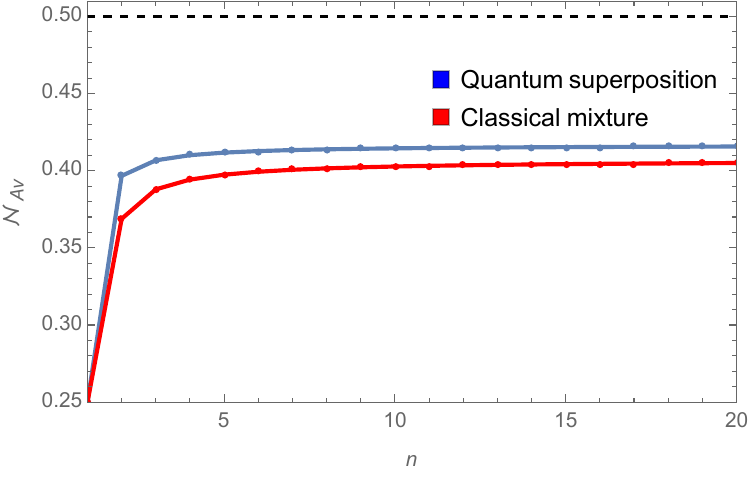}   
\caption{Average entanglement negativity, $\mathcal{N}_{\rm Av}[\rho]$, for a black hole that is in a quantum superposition of many mass states, compared to a classical mixture. Blue: the average entanglement negativity $\mathcal{N}_{\rm Av}[\rho]$ as a function of $n$, for a black hole in a superposition of $n$ states, with the squeezing parameters $r_1,...,r_n$ equally spaced. As $n$ increases, the average entanglement negativity increases. The increase in entanglement negativity does not go all the way to the case of full coherence (the dashed line), but asymptotes near 0.416. Red: the entanglement negativity $\mathcal{N}[\rho_{\rm Av}]$ as a function of $n$, for a black hole in a classical mixture of $n$ states, with the squeezing parameters $r_1,...,r_n$ equally spaced. The figure  illustrates the advantage of the quantum superposition of channels over the corresponding classical mixture.
    }\label{fig:Nstates}
 \end{figure}

We found that as $N$ increases, the decoherence decreases further. However, it asymptotically converges before hitting zero decoherence -- see Figure \ref{fig:Nstates} -- due to the fact that there is only a single shared environment. The generalization of the derivation above to the $N$-state superposition is straightforward. Instead of two squeezing parameters $r_1$ and $r_2$, we have $N$ parameters $r_1,...,r_N$. It is clear that there is no unique way of choosing the $r_1,...,r_N$ parameters.  Nevertheless, the result with two mass states in a  superposition (Figure \ref{fig:PerDiff}) demonstrated that the greatest advantage in the decoherence rate is achieved when the squeezing parameters are maximally spaced. Following this intuition, in  Figure \ref{fig:Nstates}, we plot the average negativity for maximally spaced squeezing parameters $r_1,...,r_N$.  The fact that the negativity does not asymptotically approach $1/2$, where there would be no loss of   entanglement, motivates further work into superpositions of multiple channels with correlated environments.

	\section*{Discussion}

We have shown that the possibility of quantum superpositions of black hole states with different masses can enhance the transmission of quantum information between two parties  near the black hole. This enhancement  comes from an overall reduction of the noise due to the coherent superposition of different noisy channels \cite{AharonovSupChannels,OiSupChannels,AAbergSupChannels,PhysRevA.72.012338,Abbott2020communication,Hler2019}.

	It is worth stressing that, in principle, quantum coherence in spacetime degrees of freedom  could serve as a resource for quantum communication also beyond the black hole scenario discussed in this paper. However, in many scenarios the contribution of quantum gravity is outweighed by that of other  interactions. The black hole scenario, instead, provides a setting where  quantum aspects of gravity are expected to play a crucial role. In this context, our result provides a Gedankenexperiment highlighting that the fundamental decoherence expected in a classical theory of gravity can be mitigated by quantum effects.  

 Our emphasis here on quantum gravity arises from the historic context of the question that we address. In previous work, it had been found that a black hole horizon decoheres nearby quantum systems \cite{BHDeg}. We show here that once the black hole is treated like a quantum system, i.e.\ allowed to be in a quantum superposition and to be entangled, the amount of decoherence that a near horizon system experiences significantly decreases. Moreover, when we refer to quantum gravity, we do not mean only high-energy quantum corrections to the gravitational interactions, as may happen near a black hole singularity. Instead, our formalism is along the lines of low-energy quantum gravity, as in linearized quantum gravity, which is commonly explored in previous literature at the interface between quantum information and gravity \cite{Paterek,Giacomini,Rovelli,Akil}. The gravity here is manifest through the existence of an event horizon, which causes the squeezing that gives rise to the two-mode-squeezed state \cite{Fabbri:2005mw}. The quantumness is manifest through the superposition of squeezing parameters, arising from the superposition of black hole masses (horizon surface gravities). In any classical or semi-classical gravitational theory, there is no superposition of event horizons, or of surface gravities at the horizons, and thus no superposition of  squeezing parameters.   Finally, in line with the equivalence principle, phenomena similar to the one we describe here are expected to occur when considering a superposition of accelerations.

Our Gedankenexperiment  could in principle be simulated with analog systems, in a similar way as  
 it was done for the simulation quantum field theory in curved spacetime with Bose-Einstein condensates (BECs) 
\cite{munoz2019observation,steinhauer2016observation}.  Under certain conditions, it is known that the phonons of a Bose gas can be described as effective photons propagating in a curved spacetime, with the metric being a function of the density and velocity flows of the condensate \cite{barcelo2011analogue}. An effective black hole can then be formed where the velocity flows are so strong that the phonons cannot escape, leading to phonons with a spectrum analogous to that of Hawking radiation \cite{munoz2019observation}. In addition, an analogue of the  entanglement entropy of the radiation has also been measured,  finding a  perfect match to that expected for real Hawking radiation \cite{steinhauer2016observation}. 

An analogue of the effect described in our paper could be obtained by generating  two entangled phonon flows, one that is in free fall with the velocity flows, and one that is held fixed against them. Then we would need to carry out the protocol described in Figure \ref{fig:BHSuperposition}, where the condensate is placed in a superposition of mass states. A superposition of different mass states of the condensate could then be generated by keeping the total number of particles of the Bose gas  fixed, and varying the proportion between  the number of particles in the condensate and the number of particles in the thermal cloud.    In fact, it is already thought that this mechanism generally leads to a superposition in  the number of particles in the condensate \cite{BogolubovJr}.  Finally, the entanglement between the two phonon flows would need to be measured. Although this would certainly be a challenging experiment, it could become possible in the not too distant  future, as similarly challenging experiments, such as the creation of BEC Schr\"{o}dinger cat states \cite{Howl_2019,LI2022}, are currently under way.

	An interesting direction of future research is to employ the reduction of decoherence observed in our work as a witness to provide evidence for the quantum nature of spacetime. In a quantum theory of gravity, placing a black hole in a superposition of masses should result in a superposition of event horizons and thus a superposition of Hawking channels, which we have shown can result in less decoherence than if there were instead a classical mixture of Hawking channels. The latter would be expected in a  stochastic but fundamentally classical theory of gravity, and so the level of decoherence Rob experiences in his protocol can be used to evidence the quantum nature of gravity. This is connected to {\color{purple} a} very recent work on how a particle detector coupled to a black hole in a superposition of masses can exhibit signatures of quantum-gravitational effects, such as signatory peaks dependent on the mass ratio of the superposed black hole \cite{foo2021quantum}.


%



\appendix

\section{Kraus decomposition of classical gravity channel} \label{app:Kraus}
	
	The channel \eqref{eq:xi}, which we  refer to as the Hawking channel  due to its relationship with Hawking radiation, can be written in terms of Kraus operators:
 
	\begin{align} \label{eq:kraus}
		\xi(\hat{\rho}_{\psi}) &= \mathrm{Tr}_{\shortminus k_R} \left[\hat{U}_{AR} (r) \left( \hat{\rho}_{\psi} \otimes \hat{\rho}_{\shortminus \bm{k}_R}  \right)\hat{U}^{\dagger}_{AR} (r) \right]\\
		&\equiv \sum_n \hat{M}_n \hat{\rho}_{\psi} \hat{M}^{\dagger}_n,
	\end{align}
 
	where $\hat{M}_n := \, ^{-}{\langle} n_{\shortminus \bm{k}_R} | \hat{U}_{AR} | 0_{\shortminus \bm{k}_R} \rangle^- =: I \otimes \hat{N}_n$, with $\hat{N}_n := \,^{-}{\langle} n_{\shortminus \bm{k}_R} | \hat{U}_{\bm{k}_R} | 0_{\shortminus \bm{k}_R} \rangle^-$. Due to the Pauli exclusion principle, $n$ can only be $0$ or $1$, resulting in the Kraus operators (see also \cite{banerjee2017characterization} for a derivation using the Choi-matrix representation of a quantum channel):
 
	\begin{align}
		M_0 &= \left(\begin{array}{cccc} c(r) & 0 & 0 & 0\\ 0 & 1 & 0 & 0 \\ 0 & 0 & c(r) & 0 \\ 0 & 0 & 0 & 1 \end{array} \right),\\
		M_1 &= \left(\begin{array}{cccc} 0 & 0 & 0 & 0\\ e^{-i \phi} s(r) & 0 &0&0\\0&0&0&0 \\ 0 & 0 & e^{-i \phi} s(r) & 0\end{array} \right),
	\end{align}
 
where $c(r) := \cos r$ and $s(r)=\sin r$.  On completion of the channel, the maximally-entangled state $\rho_{\psi}$  becomes:

	\begin{align} \label{eq:xirho}
		\xi(\rho_{\psi}) &= \frac{1}{2} \left(\begin{array}{cccc} c(r)^2 & 0 & 0 & c(r)\\ 0 & s(r)^2 & 0 & 0 \\ 0 & 0 & 0 & 0 \\ c(r) & 0 & 0 & 1 \end{array} \right).
	\end{align}
 
	Measuring the entanglement of this state using the negativity measure, the entanglement is found to be $\mathcal{N} [\xi(\rho_{\psi})] = c(r)^2 / 2$ \cite{PhysRevA.74.032326}.

    \section{Superposition  of Hawking channels} \label{app:SupChannels}
	
	The posteriori joint states of Alice and Rob  $\hat{\rho}^+_{AR}$ and $\hat{\rho}^-_{AR}$ after Rob measures the control are:
 
	\begin{align} 
		\hat{\rho}^+_{AR} &= 	\frac{\xi_{11}(\hat{\rho}_{\psi})  +  \xi_{22}(\hat{\rho}_{\psi}) + 	\xi_{12}(\hat{\rho}_{\psi})  +  \xi_{21} (\hat{\rho}_{\psi})}{\mathrm{Tr} [\xi_{11}(\hat{\rho}_{\psi})  +  \xi_{22}(\hat{\rho}_{\psi}) + 	\xi_{12}(\hat{\rho}_{\psi})  +  \xi_{21} (\hat{\rho}_{\psi})]},\\
		\hat{\rho}^-_{AR} &= 	\frac{\xi_{11}(\hat{\rho}_{\psi})  +  \xi_{22}(\hat{\rho}_{\psi}) - 	\xi_{12}(\hat{\rho}_{\psi})  -  \xi_{21} (\hat{\rho}_{\psi})}{\mathrm{Tr} [\xi_{11}(\hat{\rho}_{\psi})  +  \xi_{22}(\hat{\rho}_{\psi}) - 	\xi_{12}(\hat{\rho}_{\psi})  -  \xi_{21} (\hat{\rho}_{\psi})]},
	\end{align}	 
 
where, using the Kraus decomposition \eqref{eq:kraus}, we can write

\begin{align}
	\xi_{ij}(\hat{\rho}_{\psi}) &=  	\sum^{n=1}_{n=0} \hat{M}_{in} \hat{\rho}_{\psi} \hat{M}^{\dagger}_{jn},
\end{align}

with

\begin{align}
	M_{i0} &= \left(\begin{array}{cccc} c(r_i) & 0 & 0 & 0\\ 0 & 1 & 0 & 0 \\ 0 & 0 & c( r_i) & 0 \\ 0 & 0 & 0 & 1 \end{array} \right),\\
	M_{i1} &= \left(\begin{array}{cccc} 0 & 0 & 0 & 0\\ e^{-i \phi} s(r_i) & 0 &0&0\\0&0&0&0 \\ 0 & 0 & e^{-i \phi} s(r_i) & 0\end{array} \right),
\end{align}

and $i\in \{1,2\}$. The posterior states can then be written as:

	\begin{align}\nonumber
		&\hat{\rho}^+_{AR} = \\&\frac{1}{2A} \left(\begin{array}{cccc} (\sum^{i=2}_{i=1} c(r_i))^2 & 0 & 0 & 2\sum^{i=2}_{i=1} c(r_i)\\ 0 & (\sum^{i=2}_{i=1} s(r_i))^2 - 4 B & 0 & 0 \\ 0 & 0 & 0 & 0 \\ 2\sum^{i=2}_{i=1} c(r_i) & 0 & 0 & 4 \end{array} \right)
	\end{align}
 
	where:
 
	\begin{align}
		A &:= 3 + c(r_1) c(r_2) + c(\phi_1-\phi_2) s(r_1) s(r_2),\\
		B &:= s((\phi_1 - \phi_2)/2)^2 s( r_1) s(r_2),
	\end{align}
 
	and
		\begin{figure}[t]
		\includegraphics[trim={1.2cm 0 0 0},clip,width=7cm]{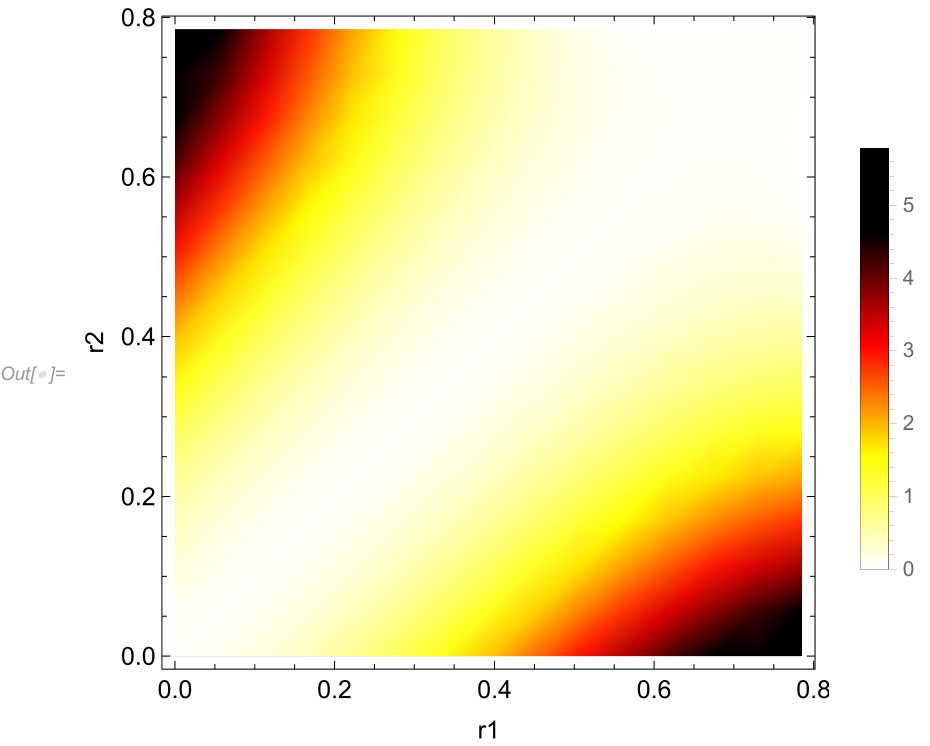}
		\caption{Density plot of percentage difference between  
        the average entanglement negativity and the classical entanglement  $\mathcal{N}^C_{Av}$. This is provided for the squeezing factors   $r_1$ and $r_2$ of the two channels for the physical range  $0$ to $\pi/4$. White through yellow and red to black indicates small to larger percentage differences.} \label{fig:PerDiffAlt}
	\end{figure}
 
	\begin{align}
		\hat{\rho}^-_{AR} &= \frac{1}{2C} \left(\begin{array}{cccc} c(r_1) - c( r_2)^2 & 0 & 0 & 0\\ 0 & (s (r_1) - s (r_2))^2 + 4 B & 0 & 0 \\ 0 & 0 & 0 & 0 \\ 0 & 0 & 0 & 0 \end{array} \right),
	\end{align}
 
	with:
 
	\begin{align}
		C := 1 - c (r_1) c (r_2) - c (\phi_1-\phi_2) s (r_1) s (r_2).
	\end{align} 
 
	Note that $\hat{\rho}^-_{AR}$ is a separable state $\frac{1}{2C}|0\rangle \langle 0| \otimes [ (c ( r_1) - c (r_2))^2 |0 \rangle \langle 0 | + ((s (r_1) - s (r_2))^2 + 4B)) |1 \rangle \langle 1 |]$. The negativity of $\hat{\rho}^+_{AR}$ when $\phi_1 = \phi_2$ is
 
	\begin{align}\nonumber
		\mathcal{N}[\hat{\rho}^+_{AR}] &= \Big[-[s( r_1) + s(r_2)]^2 \\\nonumber&+ \sqrt{16[ c(r_1) + c (r_2)]^2 + [s( r_1) + s(r_2)]^4}\Big]\\&\hspace{1cm}/4[3 + c(r_1-r_2)].
	\end{align}
 
The average entanglement that Rob would obtain after measuring the control and keeping the results of his measurements is $\mathcal{N}_{Av}[\hat{\rho}] = p_+ \mathcal{N}[\hat{\rho}^+_{AR}]$, where  $p_+$ is the probability that Rob measures $|+\rangle_c$:

\begin{align} 
	p_+ &= \mathrm{Tr} [\xi_{11}(\hat{\rho}_{\psi}) + \xi_{22}(\hat{\rho}_{\psi}) + \xi_{12}(\hat{\rho}_{\psi})+\xi_{21}(\hat{\rho}_{\psi}) ] / 4\\ \label{eq:p+}
	&= A/4.
\end{align}

In contrast, the probability that  Rob measures $|-\rangle_c$ is:

\begin{align}
	p_- &=  \mathrm{Tr} [\xi_{11}(\hat{\rho}_{\psi}) + \xi_{22}(\hat{\rho}_{\psi}) - \xi_{12}(\hat{\rho}_{\psi})-\xi_{21}(\hat{\rho}_{\psi}) ] / 4\\
	&= C/4.
\end{align}

Ignoring the phase as is conventional, or setting $\phi_1 = \phi_2$, the average entanglement $\mathcal{N}_{Av}[\hat{\rho}]:=p_+ \mathcal{N}[\hat{\rho}^+_{AR}]$ is

\begin{align}\nonumber
	\mathcal{N}_{Av}[\hat{\rho}] &= \frac{1}{16}\Big[ -[s(r_1) + s( r_2)]^2 \\&+ \sqrt{16[c(r_1) + c(r_2)]^2 + [s(r_1) + s(r_2)]^4} \Big].
\end{align} 

This is to be contrasted with the entanglement generated from an equal `classical' mixture of the channels, which has the state:

\begin{align}
	\hat{\rho}_{Av}  	&=  \frac{1}{2} \Big[	\xi_{11}(\hat{\rho}_{\psi})  +  \xi_{22}(\hat{\rho}_{\psi})\Big]\\
	&= \frac{1}{4} \left(\begin{array}{cccc} \sum^{i=2}_{i=1} c(r_i)^2  & 0 & 0 & \sum^{i=2}_{i=1} c(r_i) \\ 0 & \sum^{i=2}_{i=1} s(r_i)^2 & 0 & 0 \\ 0 & 0 & 0 & 0 \\ \sum^{i=2}_{i=1} c(r_i) & 0 & 0 & 2 \end{array} \right),
\end{align}

and negativity:

\begin{align}\nonumber
	\mathcal{N}[\hat{\rho}_{Av}] &= \frac{1}{8} \Big[-[s( r_1)^2 + s( r_2)^2]\\ &+ \sqrt{4[c(r_1) + c( r_2)]^2 +  [s( r_1)^2 + s( r_2)^2]^2} \Big].
\end{align}

This is always smaller than the average entanglement obtained from the quantum superposition of channels $\mathcal{N}[\hat{\rho}^+_{AR}]$, which  is illustrated in Figure \ref{fig:PerDiff}. 

Rather than using $\mathcal{N}[\hat{\rho}_{Av}]$, another possibility is to characterize the classical entanglement by

\begin{align}
	\mathcal{N}^C_{Av} &:= \frac{1}{2} \mathcal{N}[\hat{\rho}_1] + \frac{1}{2} \mathcal{N}[\hat{\rho}_2]\\
	&= \frac{1}{4} [c(r_1)^2 + c(r_2)^2], 
\end{align}

where $\hat{\rho}_1$ and $\hat{\rho}_2$ are the density operators obtained for the Hawking channels with $r_1$ and $r_2$. Since the negativity is not a linear function, $\mathcal{N}^C_{Av} \neq \mathcal{N}[\hat{\rho}_{Av}]$ even though  $\hat{\rho}_{Av} =  \hat{\rho}_1/2+ \hat{\rho}_2/2$. However, negativity is convex, such that $\sum_i p_i \mathcal{N}[\hat{\rho}_i] \geq  \mathcal{N}[\sum_i p_i \hat{\rho}_i]$. Despite this, the entanglement from the quantum superposition of channels is still greater than the classical mixture even by this alternative measure. That is, $\mathcal{N}_{Av}[\hat{\rho}] > \mathcal{N}^C_{Av}$ always, as illustrated by Figure \ref{fig:PerDiffAlt}.\\

	\section{Detailed analysis of quantum communication near a classical black hole} \label{app:review}
	
Here, we provide more detail on the review  of quantum communication near a classical black hole that was presented in the main text. As stated in the main text, Alice and Rob both have access to  a scalar fermionic (Grassmann) quantum field near the event horizon of a Schwarzschild black hole. For free-falling, inertial observers, the appropriate coordinates are Minkowski. Here, the quantization of the fermionic field is straightforward, resulting in particle $\hat{a}_k$, $\hat{a}^{\dagger}_{k}$ and anti-particle $\hat{b}_k$, $\hat{b}^{\dagger}_{k}$  annihilation and creation operators  that are associated with positive (particle) and negative (anti-particle) frequency,  monochromatic plane wave mode solutions in Minkowski coordinates $u^{\pm}_{\bm{k}} \propto e^{\pm i k_{\mu} x^{\mu}}$ \cite{mandl2010quantum}, where $\bm{k}$ and $k_{\mu}$ are the three- and four-momentum of the  mode, with $\mu = 1,\ldots 4$, and we are using $\eta_{\mu \nu} = \mathrm{diag}(-1,+1,+1,+1)$.  The Minkowski vacuum is defined by the absence of any mode excitations in an inertial frame $|0\rangle_M := \Pi_{\bm{k},\bm{k'}} |0_{\bm{k}}\rangle^+_M |0_{\bm{k'}} \rangle^-_M$, where $\{+,-\}$ indicate the particle and antiparticle vacua, with $\hat{a}_{\bm{k}} |0_{\bm{k}}\rangle^+_M = 0$ and $\hat{b}_{\bm{k}} |0_{\bm{k}}\rangle^-_M = 0$. 

In contrast to a free-falling observer,  a static observer just outside the event horizon can consider themselves in flat spacetime but accelerating. Mathematically this is due to the Schwarzschild spacetime metric approximating a Rindler spacetime metric near the event horizon \cite{BHDeg}, with Rindler coordinates being the natural ones for  accelerating observers in flat spacetime \cite{RindlerCoords1,RindlerCoords2}. Two different sets of Rindler coordinates are necessary for covering flat spacetime, which define two Rindler regions $\mathfrak{L}$ and $\mathfrak{R}$ that are causally disconnected from each other and admit a separate quantization procedure for a quantum field \cite{BirrelandDavies}.  We can expand a scalar fermionic field in terms of Minkowski modes $u^{\pm}_{\bm{k}}$ and, independently, in 	terms of left and right Rindler modes $u^{\pm \mathfrak{L}}_{\bm{k}}$, $u^{\pm \mathfrak{R}}_{\bm{k}}$. The Minkowski and full set of Rindler modes are, therefore, related by a change of basis and in such a way that the vacua of the two spaces differ \cite{BirrelandDavies}. This is the Unruh effect, where an accelerating observer will see an inertial vacuum filled with thermal particles \cite{Fuling73,Davies_1975,UnruhEffect,Crispino:2007eb}.

There exists an infinite number of orthonormal bases that define the Minkowski vacuum state $|0\rangle_M$. In particular, any complete set of modes made out of independent linear combinations of positive frequency Minkowski modes  will define the same vacuum. This means that we can analytically continue the left and right Rindler modes to construct positive energy Minkowski modes as linear combinations of Rindler
modes. The resulting modes are called Unruh modes $\nu^{\pm \mathfrak{L}}_{\bm{k}}$, $\nu^{\pm \mathfrak{R}}_{\bm{k}}$  that together span the  Minkowski modes and  have the  vacuum $|0\rangle_M$. The Unruh modes of Minkowski space and Rindler modes of Rindler space are simply related: the change of basis is performed by the two-mode squeezing unitary,  Eq.\ 3 in the main text. There is also an analogous expression to Eq.\ 3 for the transformation between the left Unruh operator and Rindler operators. 

Next we consider that Alice and Rob  share a  bipartite quantum state, Eq. 1 in the main text, that is written as  maximally entangled for observers in free fall:

\begin{align} 
	|\psi \rangle_{AR} =\left(|0_{\bm{k}_A}\rangle^+_A |0_{\bm{k}_R}\rangle^+_R +   |1_{\bm{k}_A} \rangle^+_A |1^{\mathfrak{R}}_{\bm{k}_R}\rangle_R^+ \right) / \sqrt{2},
\end{align}

where, as stated in the main text, $|0_{\bm{k}_A,\bm{k}_R}\rangle^+_{A,R}$ represents the Minkowski  vacuum state in modes $\bm{k}_A,\bm{k}_R$ for Alice and Rob, $|1_{\bm{k}_A}\rangle^+_A = \hat{a}^{\dagger}_{\bm{k}_A} |0\rangle^+_A$ is a one-particle Minkowski state for Alice in mode $\bm{k}_A$, and $|1^{\mathfrak{R}}_{\bm{k}_R}\rangle_R^+ \equiv \hat{a}^{R \dagger}_{\bm{k}_A} |0\rangle^+_R$ is a one-particle right Unruh state for Rob in mode $\bm{k}_R$. All other modes are assumed to be in vacua. We could have   also chosen the left Unruh state  $|1^{\mathfrak{L}}_{\bm{k}_R}\rangle_R^+$, but this  would just be equivalent to stating that Rob is then in region $\mathfrak{L}$ instead of $\mathfrak{R}$ \cite{BHDeg}. We could have also chosen an arbitrary combination of left and right Unruh modes, which is considered in \cite{SingleModeApprox}, and taking $|1_{\bm{k}_R}\rangle_R^+$ to be equivalent to a one-particle Minkowski state is referred to as the single-mode approximation \cite{PhysRevLett.91.180404,SingleModeApprox}. Here, as in \cite{BHDeg}, we do not apply such an approximation as we are interested in fundamental issues rather than a realistic experiment. We, therefore, assume that Rob has access to Unruh modes. Another possibility is that Rob builds a Minkowski wavepacket involving a superposition of general Unruh modes constructed in such way that the single-mode approximation is valid \cite{SingleModeApprox}. As we are only interested in fundamental issues, we have also only assumed simplistic global modes for Alice and Rob. We do not expect the fundamental picture to change if local modes had been used instead\footnote{See Appendix \ref{app:Local}.}.  

We now consider that Alice is free-falling just above  the event horizon, while Rob is stationary a small distance above the horizon. 	Alice, being a free-fall observer, will define her vacuum and excited states to be that of Minkowski space. Rob, however, will define his vacuum to be the Rindler (or equivalently Boulware) vacuum \cite{BHDeg}. That is, Rob will apply the transformation Eq. 3 in the main text to describe his state in the Rindler basis. However, since Rob is assumed in region I of the black hole spacetime and has no access to II, we must also trace out the Rindler modes of II that form a separate Hilbert space to that of region I. The state $|\psi \rangle_{AR}$ is thus acted upon by  the   quantum channel Eq.\ 2 in the main text, which leads to a loss of entanglement and thus also performance of quantum communication as explained in the main text.

	\section{A black hole in a superposition of masses} \label{app:BHSup}
	
	The most natural way that a black hole could be in a superposition of masses is if the original matter that collapsed to form the black hole were itself in a superposition of masses and then this superposition remains as the black hole is formed. This updates the classical no hair theorem of black holes since, in theory, the quantum phase of the superposition also characterizes the black hole.
	
	Black hole evaporation where the black hole loses mass as it emits Hawking radiation also naturally leads to an indefinite mass for the black hole since it becomes entangled with the emitted radiation \cite{Akil_2021}. 
	This is easy to see:  First, we look at the Hawking radiation state \cite{Fabbri:2005mw},

\begin{align}
\nonumber
\hspace{-0.4cm}
\ket{\rm H R } \! \! & =   \bigotimes_\omega \sqrt{1 - e^{- \frac{2\pi \omega} {\kappa }  }  } \sum_{N_ {\omega}  }  e^{-\frac{ \pi \omega} { \kappa} } 
\ket{N _ {\omega} }^{\rm int} \otimes \ket{ N_\omega} ^{\rm out} \\ \nonumber
 &\sim   \hspace{-0.5 cm} 
  \sum_{N_ {\omega_1}, N_ {\omega_2},  \dotsc }  \hspace{-0.55 cm} e^{-\frac{ \pi ( \omega_1  N_{\omega_1}+ \omega_2 N_{\omega_2}+ ...)} { \kappa} } \! \! 
\ket{N _ { \omega_1} ,N _ { \omega_2 } , ...}^{\rm int}\\ \label{Main2} &\hspace{4.5cm} \otimes 
\ket{ N_{\omega_1} , N_ {\omega_2} ,...}^{\rm out},
\end{align}
where the ``int'' and ``out'' Hilbert spaces account for the inside and the outside of the black hole respectively.
The state is a superposition of number eigenstates for each energy of the created pairs. With every burst of radiation emitted, the black hole is losing an undetermined amount of energy by eating a superposition of numbers of negative energy particles. 
For instance, say that a black hole of mass M forms after a collapse of a dust shell of mass M. Then, after the black hole emits some radiation particles, a state of the whole system will read,

 \begin{align}
  \sim   \hspace{-0.4 cm} \sum_{N_ {\omega_1}, N_ {\omega_2},  \dotsc }  \hspace{-0.3 cm} e^{-\frac{ \pi ( \omega_1  N_{\omega_1}+ \omega_2 N_{\omega_2}+ \dotsb)} { \kappa} } && \hspace{-0.3 cm} \ket{M - N_{\omega_1} \omega_1 - N_{\omega_2 } \omega_2 - \! \! \dots }_{\rm bh} \nonumber \\
 && \hspace{0.1 cm} \otimes \ket{ N_{\omega_1} , N_ {\omega_2} ,...} _{\rm out}.
 \end{align}
 
cyan Note that this follows from Eq.\eqref{Main2} based on the fact that Schwarzschild $t$ coordinate of a black hole is a Killing vector and thus energy is conserved locally and globally. Therefore the black hole mass decreases by an amount equal to that of the emitted radiation \cite{BirrelandDavies,Fabbri:2005mw}.
Therefore, the black hole will naturally go into a joint mass superposition with the Hawking particles as soon as it radiates. The Hawking particles will play the role of the control system in Eq.\ 6 in the main text. This applies to all black holes except extremal ones, which stop radiating when a particular relation between their mass and charge is satisfied \cite{Fabbri:2005mw}. The superposition will persist until the interaction between the black hole and its environment causes full decoherence.
However,  here we have ignored black hole evaporation for simplicity as in previous studies on how classical black holes lead to a degradation in quantum communication \cite{BHDeg,AliceFallsIntoBH}. 
	
	Another possibility is if Rob dropped an object that is in a superposition of energies or masses into the black hole. Since Rob is  standing above the event horizon, according to general relativity, if he were to drop a machine that is programmed to transmit information in light signals back to Rob about the distance it has covered, then Rob would in theory never receive a signal saying that the machine has actually hit the horizon, instead he forever receives signals with greater and greater wavelengths about just how close the machine is to the horizon. This is  due to the red shifting from the gravitational field. He therefore would in theory only see the machine becoming more and more smeared over the horizon and never actually dropping into it and disappearing \cite{BHsandStringRev}. However, that is assuming Rob has access to an unrealistic detector that can pick up any wavelength and can sit around for all of time. Instead, with a physical detector, at some point the detector will just stop seeing a signal since it cannot access all wavelengths, and then to all practical purposes Rob will think that the machine has fallen into the hole and will see  the event horizon appear to grow because of this. That is, the black hole's mass will appear to have increased once he no longer receives signals from the machine, which is likely to be a very short time if he is standing very close to the event horizon. This is further illustrated by considering Rob to also have access to a gravimeter which would detect the machine getting closer and closer and becoming more and more smeared over the horizon, until, due to Gauss’s law, it cannot tell the difference between the mass being smeared over the spherical horizon or being at the singularity. Therefore, dropping an object with a superposition of energies into a black hole will, to all practical purposes, place the black hole in a superposition of  masses.
	
	\section{Projective measurements of a black hole} \label{app:BhPVM}
	
	In the main text we applied a protocol such that the black hole is put into a superposition of masses  using a control qubit, and then placed in a definite mass state before a projective measurement on the control in the $\{|+\rangle, |-\rangle\}$ basis. Another option would be to leave the mass of the black hole indefinite and measure it instead in the $\{|+\rangle, |-\rangle\}$ basis. This would have the advantage of  Rob not needing to protect his state while he places the black hole in a definite mass state. In theory, it would also allow for the possibility of not needing a control, assuming that  the black hole is pre-prepared in a superposition of masses using some different means.  
	
	Without knowing the true quantum theory of gravity, it is not possible to state whether a projective measurement in the  $\{|+\rangle, |-\rangle\}$ basis exists for a quantum black hole. However, even in standard quantum theory it is notoriously difficult to perform a projective measurement to determine whether a state is in a macroscopic superposition. Assuming that a black hole is well-described by a coherent state, then a superposition of black hole masses could be described by a superposition of coherent states: $\ket{\alpha} + \ket{\beta}$. One possible method would be to first perform the displacement operator $\hat{D} (-\alpha/2-\beta/2)$ and then perform a parity operator measurement. If the measurement outcome is $+1$ then the initial state must have been $\ket{\alpha} + \ket{\beta}$. However, here the displacement operator would have to be performed very quickly since this will change the state of the black hole and thus the Hawking channel. 
	
	Assuming that the black hole can be measured in the $\{|+\rangle, |-\rangle\}$ basis, we would also need to disentangle all the modes $\bm{k} \neq \bm{k}_A, \bm{k}_R$ of the quantum field from $\bm{k}_A$ and $\bm{k}_R$. This could be achieved by, for example, Rob initially populating all the low-energy modes. Then, due to the Pauli exclusion principle, these modes would not change when Rob stands next to the superposed event horizons of the black hole, and, we can also approximate the unpopulated high-energy modes as unchanging since the strength of the Hawking channel drops off quickly with momentum.

\section{Superposition of phase channels} \label{app:SupPhases}

Above we considered a superposition of Hawking channels where there are different values of squeezing parameter $r$ but the same phase $\phi$. Usually the phase is considered to be physically irrelevant but with a quantum theory of spacetime, this may no longer be the case. Therefore, here we consider a superposition of the Hawking channels that have the same squeezing parameter but different phases. Since the phase is irrelevant in the classical mixture of channels, this also allows us to compare the resource of a superposition of channels to that of  just one channel.  

Taking a superposition of channels with the same squeezing parameter $r$ but opposite phases $\phi_1 = \phi_2 + \pi$, and assuming a similar protocol to that assumed for a superposition of squeezing parameters, the posterior states after measuring the control in a $\{|+\rangle_c,|-\rangle_c\}$ basis are

\begin{align}
	\hat{\rho}^+_{AR} &:= \frac{1}{1+\cos^2  r} \left(\begin{array}{cccc}  \cos^2 r & 0 & 0 &  \cos r\\ 0 & 0 & 0 & 0 \\ 0 & 0 & 0 & 0 \\  \cos r & 0 & 0 &1 \end{array} \right),
\end{align}

and

\begin{align}
	\hat{\rho}^-_{AR} &=  \left(\begin{array}{cccc} 0 & 0 & 0 & 0\\ 0 & 1 & 0 & 0 \\ 0 & 0 & 0 & 0 \\ 0 & 0 & 0 & 0 \end{array} \right),\\
	&= |10\rangle_{AR} \,_{AR}{\langle} 10|.
\end{align}

The state $\hat{\rho}^+_{AR}$ has entanglement:

\begin{align}
	\mathcal{N}[\hat{\rho}^+_{AR}] = |\cos r| / (1 + \cos^2 r).
\end{align}

Since we obtain zero entanglement when Rob obtains $|-\rangle_c$, the average entanglement is $p_+ \mathcal{N}[\hat{\rho}^+_{AR}]$, where $p_+$ is the same as \eqref{eq:p+} in Section \ref{app:SupChannels}, which  in this case is $p_+ = (1+\cos^2 r) / 2$ since $r:= r_1 = r_2$.  The average entanglement is then:

\begin{align}
	\mathcal{N}_{Av} [\hat{\rho}] = | \cos r| / 2.
\end{align}

\begin{figure}[t]
	\includegraphics[width=6cm]{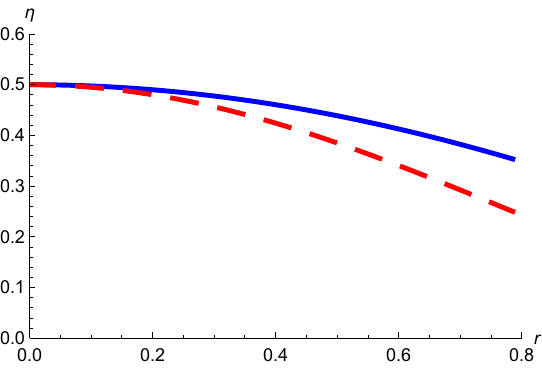}
	\caption{The blue solid line is the average Negativity of the final state of a superposition of channels  with opposite  phases $\mathcal{N}_{Av}[\rho]$ as a function of squeezing parameter $r$, while, for comparison, the red dashed line is the negativity of a single channel $\mathcal{N}[\rho] = \cos^2 r / 2$ with fixed phase.} \label{app:AvEntNegAntiPhase}
\end{figure}

This is to be compared to the entanglement obtained with the classical gravity  case of a classical mixture of the phase channels, which are equivalent, and therefore there is just one channel for which the entanglement is $\cos^2 r / 2$. The entanglement for a superposition of channels is clearly always greater than the entanglement from the classical case, as illustrated in  Figure \ref{app:AvEntNegAntiPhase}.

However, it is not clear how a superposition of phases could be achieved and what the physical meaning of this would be. Usually the phase of the squeezing operation is neglected and considered to be unphysical \cite{PhysRevA.74.032326,Audretsch1994}. One speculative possibility is that Rob could send a quantum object that has a superposition of phases into a black hole, for example the cat state $| \psi \rangle_m =  (|\alpha \rangle +  |\alpha e^{-i \pi}\rangle) / \sqrt{2 + 2 \exp(-2 \alpha^2)}$, where $|\alpha \rangle$ is a coherent state. Another option might be to consider the quantum spin of a black hole.

\section{Global mode assumption} \label{app:Local}

In the quantum communication protocol between Alice and Rob, we only considered global modes and have not considered the physical implementation of Alice and Rob's detectors. In  studies of how the classical gravitational field of a  black hole and  the acceleration of observes causes quantum decoherence,  the detectors are generally considered to be Unruh-DeWitt detectors \cite{deWittDetector}, which are point-like detectors with two internal energy levels that are  weakly coupled to a quantum matter field through a monopole interaction. Recent works have considered Unruh-DeWitt detectors in a superposition of different accelerated trajectories in Minkowski spacetime \cite{SupDeWittDetectors,PARENTANI1995227,BerryPhaseUnruhEffect,Kempf2020,Foo2020}, finding, for example, that the state of the detector excitations is not in general a convex mixture of the thermal spectrum expected for a classical mixture of accelerations. Most recently, a single particle detector in the presence of a black hole in a superposition of masses has been considered, with similar results \cite{foo2021quantum}. This is related to the work discussed here. However, in the protocol used in this work, at the end of the protocol the black hole is brought back to a definite mass state before a detector would be used. There is, therefore, no need to consider the complications induced by a superposition of Unruh-DeWitt detectors in the protocol considered in the main text. 

Localized 	wave-packets rather than global modes have been considered for investigations of entanglement degradation near classical black holes, with the same conclusions as in the  global case \cite{BHDeg}. We, therefore, expect that utilizing localized 	wave-packets instead of global modes would not entirely negate the alleviation of degradation facilitated by quantum gravity.

\section{Protecting Rob's state} \label{app:storeRho}

In the protocol in the main text Rob must protect his state while his quantum machine accesses the control to determine whether a mass should be dropped into the black hole or not. This is done so that the black hole becomes disentangled from Rob's state, and he then only needs to measure the control. The way this is achieved in the main text is to move Rob's state to a high energy mode, e.g.\ using a $90^o$ beam splitter, since the strength of the Hawking channel drops off quickly with momentum. However, there are other ways by which Rob could protect his state. For example, Rob could, in theory, store his state in helicity modes  of a spin-1 field since these modes are unaffected by the Hawking channel \cite{Ling_2007}.

\section{A black hole in a superposition of locations}

In aiding  quantum communication between Alice and Rob, we could have  used a superposition of locations of the black hole rather than black hole masses, such that Rob is at a different distance from the event horizon in both cases, although always still close to the event horizon. Exactly the same result as above would be obtained with just $r_1$ and $r_2$ following  Eq.\ 4 in the main text but with different values of $R_0$. This illustrates that the same gain would be obtained also if Rob's detectors were placed in a superposition of accelerations, without reference to quantum gravity. This, however, does not diminish the fact that quantum gravity can be used, in principle, as a resource to allay decoherence expected with classical black holes and their classical gravitational fields \cite{BHDeg,AliceFallsIntoBH}. 

There are also attempts to keep gravity classical even when the matter source is described quantum mechanically. In this case, when a black hole is put in a superposition of masses its gravitational field is not in a superposition of states as it remains a classical entity. For example, in fundamental semi-classical gravity, the gravitational field would be that generated by a black hole with a mass which is the expectation of the superposed mass state \cite{ROSENFELD1963353,moller1962theories,kibble1978relativistic,Kibble_1980}. We do not consider this possibility here.

A black hole that is  in a superposition of macroscopic locations has been considered several times before although not in the context of Hawking channels and as a resource for quantum communication (see e.g.\ \cite{SimQuantumSwitchQG,arrasmith2019decoherence}). Possible ways for the black hole to be in a superposition of locations include the collapsing matter being in a location of two positions,  and Rob possibly firing an object with a superposition of momenta into the black hole such that the back-reaction of the black hole also places it in a superposition of two momenta. Perhaps then Rob could fire another object with a superposition of momenta to make the black hole stationary in both superposition channels but now with distinct locations.

\begin{acknowledgments}
	This work is supported by the Chinese Ministry of Science and Technology (MOST) through grant 2023ZD0300600, by the Hong Kong Research Grant Council (RGC) through grants SRFS2021-7S02, 18310725, and R7035-21F, and by the State Key Laboratory of Quantum Information Technologies and Materials, Chinese University of Hong Kong. Research at Perimeter Institute is supported in part by the Government of Canada through the Department of Innovation, Science and Economic Development and by the Province of Ontario through the Ministry of Colleges and Universities. AA thanks GC for the hospitality during an early visit to the University of Hong Kong. HK acknowledges the UK Engineering and Physical Sciences Research Council (EPSRC) through grant EP/R513295/1. RH and GC acknowledge the support of grant ID 62312 from the John Templeton Foundation, as part of the QISS project, and RH acknowledges support from the John Templeton Foundation grant ID 63683, as part of the WOST project, and grant ID 62420. The opinions expressed in this publication are those of the authors and do not necessarily reflect the views of the John Templeton Foundation.
	\end{acknowledgments}






\end{document}